\documentclass{llncs}

\usepackage{amsmath,amsfonts,amssymb}

\newcommand{\ACA}{\mathcal{A}}

\newcommand{\locACA}{\delta_{\ACA}}

\newcommand{\alphA}{Q_{\ACA}}

\newcommand{\limA}{\Omega_{\ACA}}
\newcommand{\globA}{G_{\ACA}}

\newcommand{\ZZ}{\mathbb{Z}}
\newcommand{\NN}{\mathbb{N}}
\newcommand{\WW}{\mathcal{W}}

\newcommand{\cylinder}[2]{[#1]_{#2}}
\newcommand{\cyl}[1]{[#1]_0}
\newcommand{\centercyl}[1]{[#1]_{\text{mid}}}
\newcommand{\pers}[2]{\Upsilon_{#2}(#1)}
\newcommand{\langpers}[2]{L_{\Upsilon,#2}(#1)}

\title{On the Complexity of Limit Sets of Cellular Automata Associated
  with Probability Measures}

\author{Laurent
  Boyer\thanks{\email{laurent.boyer@ens-lyon.fr}}\inst{1} \and Victor
  Poupet\thanks{\email{victor.poupet@ens-lyon.fr}}\inst{1} \and
  Guillaume
  Theyssier\thanks{\email{guillaume.theyssier@univ-savoie.fr}}\inst{2}}

\institute{LIP (UMR 5668 --- CNRS, ENS Lyon, UCB Lyon, INRIA), ENS
  Lyon, 46 all\'ee d'Italie, 69364 LYON cedex 07 FRANCE \and LAMA (UMR
  5127 --- CNRS, Universit\'e de Savoie), Universit\'e de Savoie,
  Campus Scientifique, 73376 Le Bourget-du-lac cedex FRANCE}

\begin{document}

\maketitle

\begin{abstract}
  We study the notion of limit sets of cellular automata associated with
  probability measures ($\mu$-limit sets). This notion was introduced
  by P.~K{\r u}rka and A.~Maass in \cite{mulimset}. It is a refinement
  of the classical notion of $\omega$-limit sets dealing with the
  typical long term behavior of cellular automata. It focuses on the
  words which probability of appearance does not tend to $0$ as time
  tends to infinity (the persistent words).  In this paper, we give a
  characterization of the persistent language for non sensitive
  cellular automata associated with Bernouilli measures. We also study
  the computational complexity of these languages. We show that the
  persistent language can be non-recursive. But our main result is
  that the set of quasi-nilpotent cellular automata (those with a
  single configuration in their $\mu$-limit set) is neither
  recursively enumerable nor co-recursively enumerable.
\end{abstract}

\section{Introduction}

Cellular automata (CA for short) are discrete dynamical systems given
by a very simple syntactical definition. They consist of a large
collection of identical cells which evolve according to uniform local
interactions.  Despite the simplicity of the model, they are capable
of producing a wide range of different behaviors.  One of the main
challenges in the field is to give pertinent classifications of these
dynamical systems.

There has been a huge amount of attempts in the literature (see
\cite{Wolfram84,Kurka97,gpcarr}). Among them, the notion of
$\omega$-limit set has received a great interest since the results
obtained by K.~{\v C}ulik \textit{et al.}  in \cite{Culik89}. This
notion (which comes from classical dynamical systems theory) is an
attempt to catch the long term behavior of cellular automata. More
precisely, the $\omega$-limit set is the set of configurations that
may appear in the evolution after an arbitrarily long time. From a
topological point of view, it is also the largest attractor. As shown
by J.~Kari, $\omega$-limit sets can hold a great complexity since any
non-trivial property concerning them is undecidable \cite{kari94-2}.
Among such properties, the nilpotency is the simplest one: a CA is
nilpotent if its $\omega$-limit set is reduced to a single
configuration. This property is extremely strong since it implies that
all initial configurations lead to the same uniform configuration.

The major drawback of $\omega$-limit sets is that it gives the same
importance to all configurations. Thus, a negligible set of
configurations can influence the $\omega$-limit set of a CA and hide
properties of its ``typical'' behaviour.

Recently, P.K{\r u}rka and A.~Maass introduced in \cite{mulimset} a
notion of limit set associated with a probability measure ($\mu$-limit
set).  Intuitively, this notion catches the ``typical'' long term
behavior of CA. More precisely, it is defined from the patterns whose
probability of appearance doesn't go to $0$ as time goes to infinity.
So, as opposed to classical limit sets, it does not deal with what
\emph{may} appear in the long term behavior but focuses on what
\emph{does} typically appear. This difference makes the $\mu$-limit
set more suitable to study some dynamics (see \cite{mulimset}).
Moreover, it is a better tool to give theoretical justifications to
many phenomena observed experimentally (since experimentations are not
exhaustive, they must restrain to ``typical'' orbits).

In this paper, we mainly study this set from a computational
complexity point of view. We first give a new characterization of
$\mu$-limit sets associated with Bernouilli measures for any non
sensitive CA. Our characterization shows that the $\mu$-limit set does
not depend on the measure.

Then we focus on the quasi-nilpotency property: a CA is
$\mu$-quasi-nilpotent if its $\mu$-limit set is reduced to a single
configuration.  One can think that the undecidability behind limit sets
disappears as soon as we no longer consider all configuration but only
``typical'' ones. We show that this is not the case, the Turing
degree of the quasi-nilpotency problem is even higher than that of the
nilpotency problem: the set of quasi-nilpotent CA is neither
recursively enumerable nor co-recursively enumerable. The construction
used to obtain this result also allows us to show that some CA have
non recursive $\mu$-limit language.

\section{Definitions}
\label{sec:defi}

Formally, a \emph{one-dimensional CA} $\ACA$ is a triple
$(\alphA,r,\locACA)$, where $\alphA$ is a finite set of states called
the \emph{alphabet}, $r$ is the \emph{radius} and
${\locACA:\alphA^{2r+1}\rightarrow\alphA}$ is the \emph{local rule}.
A \emph{configuration} $c$ describes the state of all cells at a given
time: this is a mapping from $\ZZ$ to $\alphA$.  The set of all
possible configurations is denoted $\alphA^{\ZZ}$.  For
${c\in\alphA^\ZZ}$, we will often denote by $c_z$ the value of $c$ at
${z\in\ZZ}$.

The local description of the CA induces a global evolution.  At every
step of the computation, the configuration changes according to the
global transition rule $\globA: \alphA^\ZZ\rightarrow\alphA^\ZZ$
induced by the locale rule as follows:
\[\globA(c)_i=\locACA(c_{i-r}...c_{i}...c_{i+r}).\]

In the sequel, when considering a CA $\ACA$, we implicitly refer to
the triple $(\alphA,r,\locACA)$, where the same symbol $\ACA$ denotes
both the local and the global map

We denote by $\alphA^*=\bigcup_{n\in\NN}\alphA^n$ the set of all
finite words over $\alphA$. The \emph{length} of $u=u_1u_2...u_n$ is
$|u|=n$, and, $\forall a \in\alphA$, $|u|_a $ is the number of
occurences of $a$ in $u$. $\forall 0< i \leq j\leq |u|$, we also
define $u_{[i,j]}=u_iu_{i+1}...u_j$ and $c_{[i,j]}$ for
${c\in\alphA^\ZZ}$ in a similar way.  A word $u$ is a \emph{subword}
of a word $v$ if there exist $i$ and $j$ such that $u=v_{[i,j]}$.

For every $c\in\alphA^\ZZ$, the \emph{language of $c$}, denoted by
$L(c)$, is defined by 
\[L(c)=\{u\in\alphA^*: \exists i\in\ZZ, u=c_{[i,i+|u|-1]}\}.\] The
language of a subset of $\alphA^{\ZZ}$ is the union of the languages
of its elements.

The \emph{limit set} of a CA $\ACA$ is given by $\limA=
\bigcap_{n\in\NN}\ACA^n (\alphA^\ZZ)$. Intuitively, a configuration is
in the limit set if and only if it may appear after an arbitrarily
long evolution. A CA is said to be \emph{nilpotent} if its limit set
is reduced to a single configuration.

For every $u\in\alphA$ and $i\in\ZZ$ we define the \emph{cylinder}
$[u]_i$ as the set of configurations containing the word $u$ in
position $i$: \[[u]_i=\{c\in\alphA^{\ZZ} : c_{[i,i+|u|-1]}=u\}.\]

Let $\ACA$ be any CA and $\mu$ any Borel probability measure on
$\alphA^\ZZ$ (a measure on the Borel sets, \textit{i.e.} the smallet
$\sigma$-algebra containing open sets). For any ${n\geq 0}$,
${\ACA^n\mu}$ denotes the probability measure such that for any Borel
set ${U\subseteq \alphA^\ZZ}$ we have ${\ACA^n\mu(U) =
  \mu\bigl(\ACA^{-n}(U)\bigr)}$. If ${\alphA=\{a_1,\ldots,a_n\}}$ is
the working alphabet, a Bernouilli measure $\mu$ over $\alphA^\ZZ$ is
given by a probability vector ${(p_1,\ldots,p_n)}$ (${0\leq p_i\leq
  1}$ and ${\sum p_i = 1}$) such that, for any words $u\in
\alphA^\ast$ and any ${i\in\ZZ}$, ${\mu(\cylinder{u}{i})=\prod_{a\in
    \alphA}p_a^{|u|_a}}$.  A Bernouilli measure is \emph{complete} (or
with full support) if ${p_i\not= 0}$ for all $i$.

\begin{definition}[Persistent set]
  Let $\ACA$ be any CA and $\mu$ be a Bernouilli measure on
  $\alphA^\ZZ$. A word ${u\in \alphA^\ast}$ is a \emph{vanishing word}
  for $\ACA$ and $\mu$ if its probability to appear (in a certain
  position) after $n$ iterations tends to $0$ as $n$ grows to
  infinity. We define the set $\langpers{\ACA}{\mu}$ of
  \emph{persistent words} for $\ACA$ and $\mu$ as the complement of
  the set of vanishing words for $\ACA$ and $\mu$:
  ${u\not\in\langpers{\ACA}{\mu} \iff
    \lim_{n\rightarrow\infty}\ACA^n\mu(\cyl{u}) = 0.}$ Then the
  \emph{$\mu$-persistent set} or \emph{$\mu$-limit set} of $\ACA$ is
  the subshift $\pers{\ACA}{\mu}$ defined by ${\langpers{\ACA}{\mu}}$,
  precisely ${\pers{\ACA}{\mu} = \bigl\{c\in \alphA^\ZZ :
    L(c)\subseteq\langpers{\ACA}{\mu}\bigr\}.}$
\end{definition}

When considering limit sets, the most studied property is certainly
the nilpotency. By analogy, we may define the notion of
$\mu$-quasi-nilpotency associated with the $\mu$-limit-set.

\begin{definition}[Quasi-nilpotency]
  Let $\ACA$ be a CA and $\mu$ be any Bernouilli measure over
  $\alphA^\ZZ$. $\ACA$ is said to be \emph{$\mu$-quasi-nilpotent} if
  $\pers{\ACA}{\mu}$ is reduced to a single configuration.
\end{definition}

One can verify that a CA $\ACA$ is $\mu$-quasi-nilpotent if and only
if there is some state ${q\in \alphA}$ such that
${\langpers{\ACA}{\mu}=q^\ast}$.

\begin{definition}[Walls]
  Let $\ACA$ be any CA.  For any ${u\in \alphA^\ast}$, we denote by
  $\centercyl{u}$ the following set of configurations of $\alphA^\ZZ$:
  \[\centercyl{u} =
  \begin{cases}
    \cylinder{u}{-\frac{|u|}{2}} &\text{ if $|u|$ is even,}\\
    \cylinder{u}{-\frac{|u|+1}{2}} &\text{ if $|u|$ is odd.}
  \end{cases}\]
  A \emph{wall} for $\ACA$ is a sequence ${\WW=\bigl(w_n\bigr)_{n\geq
      0}}$ of non empty words of $\alphA^\ast$ such that:
  \begin{enumerate}
  \item ${\forall c\in\centercyl{w_0},\forall n\geq 1:\ACA^n(c)\in\centercyl{w_n}}$;
  \item the sequence ${\bigl(|w_n|\bigr)_{n\geq 0}}$ is non-increasing.
  \end{enumerate}
\end{definition}

Notice that a wall ${\WW=\bigl(w_n\bigr)_{n\geq 0}}$ is necessarily
ultimately periodic since $\centercyl{w_0}$ contains spatially
periodic configurations. The word $w_0$ is said to be the \emph{foot}
of $\WW$. A word is a \emph{foot of wall} for $\ACA$ if it is the foot
of some wall for $\ACA$. Any word in the period of the sequence $\WW$
will be called a \emph{brick} of $\WW$: formally, $w$ is a brick of
$\WW$ if there are ${p,n_0}$ such that, for all ${n\in\NN}$,
${w_{pn+n_0}=w}$.  A word $w\in \alphA^\ast$ is a \emph{brick of wall}
for $\ACA$ if it is a brick of some wall of $\ACA$.

The following well-known property relates the existence of bricks of
wall to the property of sensitivity to initial conditions (see
\cite{Kurka97} for a proof).

\begin{proposition}
  \label{prop:sensitive}
  A CA $\ACA$ of radius $r$ is sensitive to initial conditions if and
  only if it has no brick of wall of size $r$.
\end{proposition}

The key property behind that proposition is expressed by the following
easy-to-prove lemma.

\begin{lemma}
  \label{lem:wall}
  Let $\ACA$ be any CA of radius $r$.  If $w$ is the foot of a wall of $\ACA$
  having some brick of size at least $r$, then for any word
  ${u\in\alphA^\ast}$ there exists a wall of $\ACA$ whose foot is
  $wuw$ and which has bricks of size at least $|u|$.
\end{lemma}

\section{Properties of Persistent Sets}
\label{sec:prop}

It is well-known that the limit set of any CA $\ACA$ is either reduced
to a single configuration or infinite. This fact does not hold with
$\mu$-limit sets as shown by the following example. The same example
shows a CA whose persistent set does not contain any uniform
configuration (the limit set always does).  

\begin{example}
  Let $\ACA$ be the 184 CA in Wolfram's notation. That is, a two
  states ($\alphA=\{0,1\}$) one dimensionnal CA of radius 1.  Its
  local rule is given by: $\forall x\in\{0,1\}, \ACA(1,0,x)=
  \ACA(x,1,1)=1$ and $\ACA(0,0,x)=\ACA(x,1,0)=0$.  It can be seen as a
  simple model of traffic jam (see~\cite{Nagel92}).

  We first show that for the uniform Bernouilli measure $\mu_0$, the
  words $11$ and $00$ are both vanishing. It can be easily checked
  that ${u=u_0u_1...u_{2n+1} \in \ACA^{-n}(00)}$ implies that
  $u_1u_2...u_{2n+1}$ is a left factor of a well-bracketed string
  (where $0$ ``opens'' and $1$ ``closes''). As the proportion of such
  strings among all words of length $n$ tends to $0$ as $n$ grows to
  infinity, $\lim_{n\rightarrow\infty} \ACA^n\mu_0(00) = 0$ and $00$
  is not persistent. A similar argument shows that $11$ is also
  vanishing.
  
  Because for all $n$ there is at least one word of length $n$ in
  $\langpers{\ACA}{\mu_0}$, and $\langpers{\ACA}{\mu_0}$ is stable by
  subword, and $00$ and $11$ are not in $\langpers{\ACA}{\mu_0}$, we
  have $\langpers{184}{\mu}=(0+\epsilon) (10)^*(1+\epsilon)$ and
  $\pers{184}{\mu}=\{{}^\omega(01)^{\omega},{}^\omega(10)^{\omega}\}$.
  \qed
\end{example}
 
We will now give a characterization of the persistent language of non
sensitive cellular automata. Before stating the theorem, we need a
lemma expressing that for infinitely many steps the preimages of a
persistent word must contain any given word at some fixed position.

\begin{lemma}
  \label{euhene}
  Let $\ACA$ be any CA of radius $r$ and $\mu$ be any complete
  Bernouilli measure over $\alphA^\ZZ$. Then, for any ${w\in
    \alphA^\ast}$ and ${u\in\langpers{\ACA}{\mu}}$ there are
  positive integers $k_1$ and $k_2$ and a strictly increasing sequence
  of positive integers ${\bigl(n_j\bigr)_{j\geq 0}}$ such that
  \[\forall j\geq 0 : \ACA^{-n_j}(u)\cap\bigl(\alphA^{rn_j-k_1-|w|}\cdot
  \{w\}\cdot \alphA^{k_1+k_2+|u|}\cdot \{w\}\cdot
  \alphA^{rn_j-k_2-|w|}\bigr)\not=\emptyset.\]
\end{lemma}
\begin{proof}
  Suppose by contradiction that ${u\in\langpers{\ACA}{\mu}}$ does not
  verify the lemma. Then we have ${\forall k\geq 0, \exists n_k\geq 0, \forall n\geq n_k}$:
  \[\ACA^{-n}(u)\subseteq
  \alphA^{n-k|w|}(\alphA^{|w|}\setminus\{w\})^k\alphA^{|u|}(\alphA^{|w|}\setminus\{w\})^k\alphA^{n-k|w|}.\]
  Then, for any $k$ and any $n\geq n_k$, we have :
  ${\ACA^n\mu(\cyl{u}) \leq \bigl(1-\mu(\cyl{w})\bigr)^{2k}}$.  Thus,
  ${\ACA^n\mu(u)\rightarrow 0}$ as ${n\rightarrow\infty}$ and
  ${u\not\in\langpers{\ACA}{\mu}}$.\qed
\end{proof}

\begin{theorem}
  \label{theo:nosensitive} Let $\ACA$ be a CA which is not sensitive to
  initial conditions and $\mu$ any complete Bernouilli measure.
  Then $\langpers{\ACA}{\mu}$ is exactly the set of bricks of wall for
  $\ACA$.
\end{theorem}
\begin{proof}
  First, consider a brick of wall $u$ for $\ACA$. By definition, there
  exists a word ${w\in\alphA^\ast}$ and positive integers $n_0$ and $p$
  such that ${\forall c\in\centercyl{w}}$ and ${\forall n\geq 0}$:
  ${\ACA^{np+n_0}(c)\in\centercyl{u}}$. Thus
  ${\ACA^{np+n_0}\mu(\cyl{u})\geq \mu(\cyl{w})}$ which proves
  ${u\in\langpers{\ACA}{\mu}}$.

  Conversely, let ${u\in\langpers{\ACA}{\mu}}$. By
  proposition~\ref{prop:sensitive}, if $\ACA$ is not sensitive to
  initial conditions, it has a brick of wall of size at least $r$
  (where $r$ is the radius of $\ACA$) associated with some wall
  ${\WW=\bigl(w_n\bigr)_{n\geq 0}}$. Applying lemma~\ref{euhene} to
  $w_0$, we know there exist positive integers $k_1$ and $k_2$ and a
  strictly increasing sequence of positive integers
  ${\bigl(n_j\bigr)_{j\geq 0}}$ such that
  \[\forall j\geq 0 : \ACA^{-n_j}(u)\cap\bigl(\alphA^{rn_j-k_1-|w_0|}\cdot
  \{w_0\}\cdot \alphA^{k_1+k_2+|u|}\cdot \{w_0\}\cdot
  \alphA^{rn_j-k_2-|w_0|}\bigr)\not=\emptyset.\]

  Since $\alphA^{k_1+k_2+|u|}$ is finite, we can extract from
  ${\bigl(n_j\bigr)_{j\geq 0}}$ a sub-sequence
  ${\bigl(n_{j_k}\bigr)_{k\geq 0}}$ such that for some
  ${v\in\alphA^{k_1+k_2+|u|}}$ we have:
  \[\forall k\geq 0 : \ACA^{-n_j}(u)\cap\bigl(\alphA^{rn_{j_k}-k_1-|w_0|}\cdot
  \{w_0\}\cdot v\cdot \{w_0\}\cdot
  \alphA^{rn_{j_k}-k_2-|w_0|}\bigr)\not=\emptyset.\] By
  lemma~\ref{lem:wall}, $w_0vw_0$ is the foot of a wall of $\ACA$ with
  a brick of size at least $|v|$. By the above property, we conclude
  that $u$ is a subword of such a brick of wall.  Therefore $u$ is
  itself a brick of wall of $\ACA$.  \qed
\end{proof}



Notice that theorem~\ref{theo:nosensitive} implies that, for any CA
$\ACA$ which is not sensitive to initial conditions, the set
$\pers{\ACA}{\mu}$ is the same for any complete Bernouilli measure.


However, there exists some sensitive CA whose $\mu$-persistent set does
depend on the Bernouilli measure $\mu$ as pointed out by A.~Maass and
P.~K\r{u}rka in \cite{mulimset}: for instance the ``just gliders'' CA
$\ACA$ is sensitive to initial conditions and such that, for any
Bernouilli measure $\mu$, $\pers{\ACA}{\mu}$ is reduced to a single
configuration if and only if $\mu$ gives the same probability to two
peculiar letters of $\alphA$.

\section{Undecidability Results}
\label{sec:undeci}

This section addresses different decision problems associated with the
persistent language of cellular automata. To simplify the statement of
the studied problems, we will only consider the uniform measure. Thus,
$\mu$ will always denote the uniform measure in this section (the
working alphabet will be determined by the context). However, all the
results extend to complete Bernouilli measures using
lemma~\ref{euhene} and theorem~\ref{theo:nosensitive} from previous
section.

\begin{remark}
  In his proof of undecidability of nilpotency \cite{kari92}, J.~Kari
  actually shows that it is undecidable to determine whether a given
  CA $\ACA$ with a spreading state (a state $s$ such that
  ${\locACA(a_1,\ldots,a_n) = s}$ whenever ${s\in\{a_1,\ldots,
    a_n\}}$) is nilpotent. Moreover, it follows from
  theorem~\ref{theo:nosensitive} that such a CA is
  $\mu$-quasi-nilpotent for any Bernouilli measure $\mu$ (since the
  only bricks of wall are the words $s^n$, ${n\in\NN}$). Thus, it is
  undecidable to determine whether a $\mu$-quasi-nilpotent CA is
  nilpotent.\qed
\end{remark}

\begin{theorem}
  \label{theo:undeci}
  The set of $\mu$-quasi-nilpotent CA is not recursively enumerable.
\end{theorem}

\begin{proof}

  \newcommand{\Sim}{S_{\operatorname{simul}}}
  \newcommand{\Sig}{S_{\operatorname{signals}}} Given a Turing machine
  $M$ of states $Q_M$ and tape alphabet $\Sigma=\{0,1,B\}$ working on
  a semi-infinite tape, we will construct a CA $\ACA$ of radius 1 that
  will be quasi-nilpotent if and only if $M$ doesn't halt on the empty
  input.
   
  The states of $\ACA$ will be $\{\#\}\cup (\Sim\times\Sig)$ where
  $\#$ is an inalterable state, meaning that if a cell is in this
  state it will never change to any other state,
  $\Sim=(Q_M\cup\{-\})\times\Sigma$ is the set of states needed to
  simulate the behavior of $M$ (a state $(-,\alpha)$ represents a cell
  of the tape containing the letter $\alpha$ without the head and a
  state $(q,\alpha)$ represents that the head is on this cell in state
  $q$) and $\Sig=\{-,L,F,R,D\}$ is a set of \emph{signals} whose
  meaning and behavior will be explained later.
   
  The transition rule of the automaton can be described by the
  following rules:
  \begin{itemize}
  \item As said earlier, $\#$ states are inalterable. Since the
    automaton is of radius 1, they act as delimiters or walls, no
    information can go across them. A finite set of contiguous non
    $\#$ cells between two $\#$ states will be referred to as a
    segment. The length of the segment will be the number of cells
    between the two $\#$ states.
  \item At all times, all cells not in the $\#$ state will simulate
    the behavior of $M$ on their first component. We deal with
    conflicts (two heads that want to move on a given cell for
    example) in any given way, since we'll see that these have no
    impact on what we'll do later (ultimately, we'll only be
    interested in regular simulations starting on an empty input). If
    at some point in the computation the head wants to move to a cell
    in state $\#$, the head is deleted so that the computation cannot
    end.
  \item The signal $-$ means that there is in fact no particular
    signal on the cell.
  \item If at some point in the computation the final state $q_f$ of
    $M$ is reached, the cell where this state appears generates a
    signal $F$ (on its ``signal'' component).
  \item The $F$ signal moves towards the left at maximum speed. When
    it reaches the left border of the segment ($\#$) it turns into an
    $R$ signal.
  \item The $R$ signal will move to the right and while doing so it
    will reset the computation that is held on the first component of
    the cells it moves through, meaning that it will put the head in
    its initial state $q_0$ on the first cell of the segment and put a
    blank symbol $B$ on every cell of the tape. Since this signal
    moves at maximum speed, the simulation of $M$ can occur without
    problems on a clean tape.
  \item When the $R$ signal meets the right end of the segment it
    disappears.
  \item During all this time, the rightmost cell of a segment (any
    cell that is on the left of a $\#$ cell) will generate $L$ signals
    at every time.
  \item $L$ signals move to the left at maximum speed. When one of
    these signals reaches the left border of a segment, it generates a
    $D$ signal.
  \item The $D$ signals destroy the whole segment by moving to the
    right while changing all the cells they go through into $\#$.
    They obviously disappear when they meet a $\#$ cell since they
    can't go any further.
  \end{itemize}
   
  All the signals that we use move at maximum speed (one cell per
  step) in one of the two available directions. Signals going in
  opposite directions are not allowed to cross each other, thus, one of
  the two must disappear. The priority is as follows:
  \[
  L<F<R<D
  \]

  For example, if an $R$ signal is moving to the right (while cleaning
  the computation) and an $L$ signal is moving to the left, when they
  meet, the $R$ signal keeps moving to the right and the $L$ signal
  disappears.
   
  Let's assume that $M$ halts in $t$ steps and let's consider the
  segment of length $2t$ in which there are no signals on any cell,
  the first cell is in state $(q_0,B)$ and all other cells are in
  state $(-,B)$. On this segment, the simulation of $M$ starts from a
  well formed configuration so it will reach the $q_f$ state after $t$
  steps and generate an $F$ signal.  Meanwhile $L$ signals appear from
  the right border and move to the left.  Because the segment is of
  length $2t$, the $F$ signal appears on the left of all $L$ signals,
  so it reaches the origin before all $L$ signals and creates an $R$
  signal. This $R$ signal will reset the computation while deleting
  all $L$ signals. From there a new computation starts that will have
  enough time to finish again and delete the $L$ signals again.
  Because the segment is ``protected'' from any outside interference
  by the $\#$ cells, this cycle will continue forever and no $\#$
  state will appear on the segment. Because there are only a finite
  number of possible configurations on the segment the automaton
  eventually enters a cycle on this non-empty segment. According to
  theorem \ref{theo:nosensitive} this segment is part of the persistent
  language so $\ACA$ is not $\mu$-quasi-nilpotent.
  
  Now we will assume that $M$ doesn't halt and show that any segment
  of length $n$ disappears after at most $5n$ steps. The proof is
  based on the observation that we can't delay the apparition of a $D$
  signal on the first cell of the segment for more than $4n$ steps.
   
  It's possible that there was already a $D$ signal somewhere on the
  segment in the inital configuration. In this case, the $D$ signal
  will cut the segment in two by creating a $\#$ state where it was
  initially and then delete the right part of the segment. This means
  that if there is a $D$ signal on a segment in the initial
  configuration we can focus on a shorter segment on which there is no
  $D$ initially and let the already present $D$ take care of the rest
  of the segment.
   
  Therefore we can assume that the segment we are studying doesn't
  contain any $D$ initially. This means that after at most $n$ steps
  all original $R$ signals will have disappeared. From there, $L$
  signals will start appearing on the right border of the segment and
  try to proceed to the left (they would arrive at time $2n$). To stop
  them from reaching the left border and generating a $D$ signal, the
  only possibility is to generate an $R$ signal on the left border of
  the segment before the time $2n$. From there, the $R$ signal will
  reset the configuration of the simulation so that what is computed
  on the left of this $R$ signal is a normal computation of $M$ on the
  empty input. Since we have assumed that $M$ doesn't halt, this
  ``well formed'' computation will not reach the $q_f$ state. When the
  $R$ signal reaches the right end of the segment (at time at most
  $3n$), it disappears and the $L$ signals start moving to the left
  again. Since the simulation of $M$ doesn't reach the final state no
  $F$ signal is generated so there's nothing to stop the $L$ signals
  from reaching the left border, generate a $D$ signal and delete the
  whole segment. The whole segment is therefore deleted after at most
  $5n$ steps.

  To complete the proof, we need only show that in this case (if $M$
  doesn't halt) no other state than $\#$ can appear in a brick of
  wall. Let's consider a wall $\WW=(w_i)_{i\in\NN}$. Let's consider
  the configuration $c_{w_0}$ containing $\#$ states everywhere except
  on its center where it is the word $w_0$. Obviously $c_{w_0}$ is in
  $\centercyl{w_0}$ and doesn't contain any segment longer than
  $|w_0|$ so no segment will survive more than $5|w_0|$ steps, which
  means that for any $n\geq 5|w_0|$, $\ACA^n(c_{w_0})$ is the uniform
  $\#$ configuration, which implies that $w_n\in\#^*$. From theorem
  \ref{theo:nosensitive} we conclude that $\ACA$ is
  $\mu$-quasi-nilpotent.  \qed
\end{proof}

\begin{corollary}
  Given a CA $\ACA$ and a word $w$, the property that $w$ is not
  persistent for $\ACA$ is not semi-decidable. In other words the set
  $\{(\ACA,w) | w\notin\langpers{\ACA}{\mu}\}$ is not recursively
  enumerable.
\end{corollary}

\begin{proof}
  We know that a CA is quasi-nilpotent if and only if only one of its
  states is persistent.  If we could semi-decide that a given state is
  not persistent, then we could use this algorithm on all states in
  parallel and if the CA is quasi-nilpotent the algorithm would
  eventually show that all but one states are not persistent, thus
  showing that the CA is quasi-nilpotent. We would therefore have an
  algorithm to semi-decide that a CA is quasi-nilpotent, which is in
  contradiction with theorem \ref{theo:undeci}.\qed
\end{proof}

\begin{remark}
  The proof above shows that it is also undecidable to determine
  whether the persistent set is finite or not. Indeed, it is not
  difficult to check that the persitent set of the constructed CA is
  either reduced to a single configuration or infinite.\qed
\end{remark}

\begin{theorem}
  There exists a CA with a non-recursive persistent language.
\end{theorem}
\begin{proof}[sketch]
  It is possible to show this by slightly modifying the CA constructed
  in the proof of theorem \ref{theo:undeci}. To do so we use another
  layer in the states so that each regular cell of a segment also has
  a ``memory'' containing a tape symbol. The memory of a cell can
  never be changed (except when the cell becomes $\#$ in which case
  the memory is lost). Instead of starting from an empty input when
  the simulation of $M$ is reset by an $R$ signal it's the memory of
  each cell that's written on the tape. This way we can simulate the
  behavior of $M$ on any input. It is then easy to prove that a
  segment survives if and only if the memory of its cells corresponds
  to a word $wB^k$ where $M(w)$ ends using less than $|w|+k$ cells.

  If the persistent language of $\ACA$ is recursive, then the language
  $\#wB$ such that $M(w)$ halts is also recursive: there is a segment
  in the persistent language whose memory layer is $wB^k$, so there is
  a word of memory $\#wB$ (stability by factor).  Therefore if the
  domain of $M$ is not recursive (a universal machine for instance)
  neither is $\langpers{\ACA}{\mu_0}$.
\end{proof}

\begin{theorem}
  The set of $\mu$-quasi-nilpotent CA is not co-recursively
  enumerable.
\end{theorem}

\begin{proof}
  \newcommand{\Sig}{S_{\operatorname{signals}}} As with the proof of
  theorem \ref{theo:undeci}, we will consider a Turing machine $M$ and
  create a cellular automaton $\ACA$ of radius $1$ that simulates $M$.
  $\ACA$ will be quasi-nilpotent if and only if $M$ halts on the empty
  input. As earlier, the configuration will be divided in
  \emph{segments} separated by $\#$ cells, however in this case $\#$
  states won't be totally inalterable. The idea is that we will again
  simulate the behavior of $M$ on each segment but now if the
  simulation doesn't halt the right $\#$ of the segment will be erased
  so that the available space for the simulation is increased, and the
  simulation will start again. If at some point the simulation ends
  then the segment is erased. This way non-empty segment will remain
  on the configuration if the machine $M$ doesn't halt but almost
  every segment will be deleted if the machine halts.

  The construction will be very similar to the previous one (proof of
  theorem \ref{theo:undeci}). The states of $\ACA$ are almost the same,
  the new set of signals being $\Sig=\{-,L,R,D,D_L,D_R,C_L,C_R\}$.
  
  The evolution of the automaton is described as follows:
  \begin{itemize}
  \item The $\#$ state is now ``almost'' inalterable in the sense that
    only one particular signal ($D$) can erase it. We will continue to
    use the notion of segment (finite set of contiguous cells between
    two $\#$).
  \item The simulation of $M$ takes place on each segment as in the
    previous construction.
  \item $L$ signals will appear continuously on the right border of a
    segment and proceed to the left.
  \item When an $L$ signal meets the $\#$ cell at the left border of
    the segment it turns into a $D$ signal.
  \item $D$ signals move to the right. They erase all $L$ signals they
    meet. If a $D$ signal finds a final state $q_f$ in the simulation
    of $M$, it generates two signals $D_L$ and $D_R$ that will erase
    the segment (turn all cells into $\#$) by propagating to the left
    and right respectively until they reach the end of the segment.
    If the $D$ signal doesn't see any $q_f$ state and reaches the
    right $\#$ of the segment it turns it into a regular cell and
    creates two signals $C_L$ and $C_R$ on the cells next to where the
    $\#$ cell was.
  \item The $C_L$ and $C_R$ signals move to the left and to the right
    respectively. Their function is to clear the segment so that a
    fresh simulation of $M$ can start back from the beginning. Both
    signals will erase any signal they come across.  When the $C_L$
    signal reaches the beginning of the segment it turns into an $R$
    signal. When the $C_R$ signal reaches the end of the segment it
    disappears.
  \item The $R$ signal moves to the right and resets the simulation as
    it moves as in the previous proof. It also erases all $L$ signals.
  \end{itemize}
  
  Proving the theorem from this construction will now be similar to
  the proof of theorem \ref{theo:undeci}. The $\#$ states can only be
  deleted by a signal that comes from their left so if two segments
  merge it's because the merging signal came from the leftmost of the
  two segments while the rightmost one can do nothing to prevent it.
  We'll say that the the left segment \emph{invades} the right one.

  Let's see what happens if $M$ doesn't halt on empty input. In that
  case a ``normal'' simulation of $M$ will never reach the $q_f$ state
  so the $D_L$ and $D_R$ signals should never appear. On any segment
  where there are initially no signals and no simulation of $M$ going
  on $L$ signals will appear, reach the left border, and generate an
  $R$ signal that will start a new correct simulation. This simulation
  will not end so the segment will eventually invade the one on its
  right and when doing so $C_L$ and $C_R$ signals will appear to clean
  the segment and a new correct simulation will again take place on
  the wider segment, etc. Since no matter how wide the segment is the
  simulation will never end the segment will never disappear. It is
  also possible that the segment we have considered is eventually
  invaded but when the invasion occurs $C_L$ and $C_R$ signals appear
  that will clean the wider segment and ensure that the new simulation
  that takes place on this segment is also correct so again there's no
  risk that the segment disappears.

  In other words, if $M$ doesn't halt, any ``inactive'' segment on the
  initial configuration will grow and survive forever (the cells that
  were initially on this segment will never become $\#$). Let $s$ be
  such an ``inactive'' segment of length $2k+1$ including the border
  $\#$, then for all $n\in\NN$ and all $w\in\alphA^n$,
  \[wsw\in\bigcup_{q\in\alphA\setminus\{\#\}} A^{-n-k}(q)\] 

  This means that $\sum_{q\in\alphA\setminus\{\#\}}\ACA^n\mu(q)\geq
  \mu(\cyl{s})$ so at least one of the non-$\#$ states is persistent.
  
  Now we have to check that if the machine $M$ halts in $t$ steps then
  no other state than $\#$ is persistent. Let's consider a segment $s$
  of length $l\geq 2t$ on which there is no simulation of $M$ going on
  and the only signal present is an $R$ signal on the first cell.
  While this segment is not invaded, it will simply do correct
  simulations of $M$, reach the final state in time so that the $D$
  signal sees it, the $D_L$ and $D_R$ signals will therefore appear
  and turn the whole segment into $\#$. This means that such a segment
  doesn't invade its right neighbor. Moreover, if such it happens to
  be invaded by its left neighbor, the invasion will make $C_L$ and
  $C_R$ segments appear, which will ensure that on this new segment a
  new correct simulation starts. The segment will do correct
  simulations and grow until it's big enough so that a simulation
  ends. This will happen before all the $\#$ from $s$'s disappearance
  have been deleted because there will be enough room to complete a
  simulation before that so this other segment will also turn to $\#$
  before going past the initial boudaries of $s$. This means that if
  $M$ terminates there exists a segment such that no matter what
  happens it will never invade its right neighbor. We'll call such a
  segment \emph{non-invasive}.
  
  Let's see what happens to a segment such that there is a
  non-invasive segment on its left at a distance $d_l$ (the distance
  is taken from the right border of the non-invasive segment to the
  left border of the considered segment) and a $\#$ on its right at a
  distance $d_r\geq 2t$ (taken from the right border of the segment).
  
  While the segment is not invaded it will after some time that we
  can bound easily depending on its length start a correct simulation
  or be completely deleted (because the $L$ signals cannot be delayed
  forever). From there it will continuously do simulations and invade
  its neighbors if the simulations do not halt. Since there is a $\#$
  at a distance $d_2\geq 2t$, the simulation will eventually end
  before this $\#$ symbol is deleted since the segment will be wide
  enough, so the segment will eventually disappear. The only thing
  that could delay the disappearance of the segment would be a series
  of invasion of the segment. However, since there is a non-invasive
  segment on the left of segment, we know that there is only a limited
  number of possible invasions so we can bound the time until all
  possible invasions have occurred. From there, the simulation will
  start correctly on a segment that will not be invaded and will
  therefore disappear.
  
  To sum up, we have shown that if $M$ halts, there exists a function
  $\sigma:\NN^2\rightarrow\NN$ such that any segment that has a
  non-invasive segment on its left at a distance $d_1$ and a $\#$
  cells on its right at a distance $d_2\geq 2t$, will disappear after
  at most $\sigma(d_1,d_2)$ steps. This means that for any $n\geq
  \sigma(d_1,d_2)$ and any $q\in\alphA\setminus\{\#\}$, any word in
  $\ACA^{-n}(q)$ has no non-invasive segment on the cells left of the
  position $-d_1$ and no two $\#$ symbols on the cells between
  positions $2t$ and $d_2$ (the first one is the end of the segment,
  that can possibly be deleted by an already-present $D$ signal). This
  restriction implies that none of these states is persistent (see
  lemma~\ref{euhene}).
  \qed
\end{proof}

\begin{corollary}
  Given a CA $\ACA$ and a word $w$, the property that $w$ is
  persistent for $\ACA$ is not semi-decidable. In other words the set
  $\{(\ACA,w) | w\in\langpers{\ACA}{\mu}\}$ is not recursively enumerable.
\end{corollary}

\section{Conclusion and Perspectives}

We proved that the $\mu$-quasi-nilpotency property is neither
recursiveley enumerable nor co-recursively enumerable. In our opinion,
such a result has two interesting aspects. First, it deals with a kind
of problem rarely considered in the literature: a property of
``typical'' or random configurations only. We believe that such
properties are closer to what experimental observations may capture and
therefore that our undecidability results have a stronger meaning to
physicists or other scientists concerned with modelisation using
cellular automata.  Second, it gives an example of a ``natural''
property of cellular automata with a high Turing degree (few examples
are known, see~\cite{Sutner03c}).

A natural way to continue the study of the computational complexity of
persistent sets would be to try to prove a Rice theorem for
$\mu$-limit sets. Any property concerning limit sets is either trivial
or undecidable. Is it the same for $\mu$-limit sets?

Another interesting research direction would be to understand better 
how the probability of appearance of some word can vary with time.
More precisely, we left open a very simple question: do we have
$\langpers{\ACA}{\mu}=\langpers{\ACA^t}{\mu}$ for any CA $\ACA$ and
any $t$ ?

Finally, we can also consider extensions of our work to a
broader class of measures or by raising the dimension. In the latter
case, the notion of wall does not play the same role (a finite pattern
does not cut a bi-dimensional configuration into two disconnected
components) and the case of non-sensitive CA is to be reconsidered.

\bibliographystyle{splncs}
\bibliography{ac}

\end{document}